# DISTRIBUTED LOG ANALYSIS ON THE CLOUD USING MapReduce

*Galip Aydin, Ibrahim R. Hallac*

Original scientific paper

In this paper we describe our work on designing a web based, distributed data analysis system based on the popular MapReduce framework deployed on a small cloud; developed specifically for analyzing web server logs. The log analysis system consists of several cluster nodes, it splits the large log files on a distributed file system and quickly processes them using MapReduce programming model. The cluster is created using an open source cloud infrastructure, which allows us to easily expand the computational power by adding new nodes. This gives us the ability to automatically resize the cluster according to the data analysis requirements. We implemented MapReduce programs for basic log analysis needs like frequency analysis, error detection, busy hour detection etc. as well as more complex analyses which require running several jobs. The system can automatically identify and analyze several web server log types such as Apache, IIS, Squid etc. We use open source projects for creating the cloud infrastructure and running MapReduce jobs.

**Distribuirana analiza zapisa na oblaku primjenom MapReduce**

Izvorni znanstveni članak

U ovom članku opisujemo naš rad na projektiranju na mreži zasnovanog sustava analize distribuiranih podataka koji se zasniva na popularnom MapReduce okviru postavljenom na malom oblaku i razvijenom specijalno za analizu zapisa web poslužnika. Sustav analize zapisa sastoji se od nekoliko čvorova klastera, dijeli velike datoteke zapisa na distribuirani sustav datoteke i brzo ih obrađuje primjenom MapReduce modela programiranja. Klaster se stvara primjenom open source infrastrukture oblaka, čime nam je omogućeno jednostavno povećanje računalne snage dodavanjem dvaju čvorova. Time nam je data mogućnost da jednostavno promijenimo veličinu klastera u skladu s potrebama analize podataka. Primijenili smo MapReduce programe za potrebe osnovne analize zapisa poput frekvencijske analize, otkrivanja greške, otkrivanja prometnog sata (busy hour) itd. kao i za složenije analize za koje je potrebno nekoliko poslova. Sustav može automatski prepoznati i analizirati više vrsta zapisa web poslužnika kao što su Apache, IIS, Squid itd. Primijenjujemo open source projekte za kreiranje infrastrukture oblaka i obavljanje MapReduce poslova.

Ključne riječi: *analiza zapisa; Hadoop; MapReduce; računarstvo u oblaku*

## 1 Introduction

In recent years Big Data analysis has become one of the most popular frontiers in IT world and keeps drawing more interest from the academia and the industry alike. The sheer amount of data generated from web, sensors, satellites and many other sources overcome the traditional data analysis approaches, which pave the way for new types of programming models such as MapReduce.

These new approaches, which are generally called frameworks, are gaining importance as the volume of data increases. High performance in processing big data is not the only necessary ability, scalability is also very important. Because each big data problem needs different amount of computing resources, this nonlinearity can be solved by the environment's scalable structure. Divyakant and his friends have discussed the importance of scalability, elasticity and other features of big data solutions in their work [1].

Web servers record the web site's user interactions on hard drives as log files usually in plain text. Web server logs generally contain information like clientID, timestamp, objectID, size, method, status, type, server etc. Log files can give an idea about which pages are requested most, the busiest time of the server etc. There are specific tools for analyzing server logs but it is not a very convenient way to keep the server busy if the processes take too much time when we have increasing gigabytes or terabytes of data.

There are also some servers coming with pre-installed analyzing tools like AWStats or Webalizer. However in general, log analysis tools cannot process very large files effectively without using high amount of resources. With MapReduce approach terabytes of log data can be fast processed by distributing the files over a cluster of commodity computers.

MapReduce based analysis systems have been proposed for similar use such as in [2] Internet traffic has been analyzed. The system is able to analyze large volume Internet flow data and MapReduce can improve flow statistics computation by 72 %. Similarly in [3] Hadoop has been used for multi-terabytes of Internet traffic analysis in a scalable manner. In [4] Hadoop has been used to detect anomalies in high volume data traffic logs. Several other studies also focus on traffic analysis or specific types of log analysis. However in this paper a general-purpose web server log analysis system, which can support several different types of web server logs, is described.

In this paper we propose a log analysis platform based on open source Apache Hadoop framework. The platform can analyze several types of web server log files in parallel. We use OpenStack, an open source Cloud Computing system for creating virtual machines to deploy Hadoop clusters.

The paper is organized as follows. In Section 2 we first present the MapReduce programming model and its open source implementation Hadoop. Then the general architecture of the proposed system is given. In section 2.1, we describe our approach for simple log analysis with a MapReduce job consisting only of one map and one reduce operation. In section 2.2, we explain how complex log analysis can be carried out using chained MapReduce jobs. In Section 3, we give the experimental results and performance evaluation of the system using Hadoop. In Section 4, performance comparison for Hadoop and Spark is given. Finally in Section 5, conclusion of this paper is presented.

## 2 Log Analysis Method

Google's MapReduce programming model provides the ability to analyze big data by using commodity hardware in a scalable manner. Processing large amounts of data with this model is not only cost effective but also simple and fault tolerant by its concept [5].

The tasks in MapReduce paradigm are mainly separated in two modules as mappers and reducers. These programs which are usually called as jobs work are as follows:

Map (k1,v1) → list (k2,v2)
Reduce (k2, list (v2)) → list (v3)

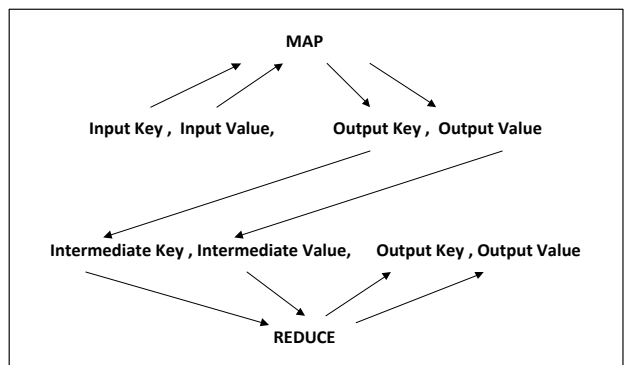

**Figure 1** Map and reduce steps

The illustration in Fig. 1 shows overall steps for a MapReduce job.

In this study a log analysis system was developed for analyzing big sets of log data using MapReduce approach. We first create a private cloud using OpenStack and build a Hadoop cluster on top of it. Apache access logs, Apache error logs, Squid logs and IIS server logs were obtained from the IT Department of Firat University.

Several MapReduce programs were written for obtaining information from log files. These programs will be explained in the next section of the paper.

Log files are non-relational, unstructured data. But they have a pre-determined format so that information about server activities can be extracted by processing them.

In this project we used the open source MapReduce execution platform called Hadoop [6]. Hadoop framework enables storing and parallel processing of large data sets. File system of Hadoop's distributed processing framework is The Hadoop Distributed File System (HDFS) which is an open source implementation of Google File System [7]. The main idea of HDFS is to split the input data into blocks and store them horizontally with their replicas on the cluster. The size of the blocks and number of the replicas are predefined in the framework configuration. A general view of HDFS is shown in Fig. 2. Here b1, b2, b3 … are the split blocks of the input data set which are stored on node1, node2, node3 … nodes in the Hadoop cluster. Replications of the blocks are stored in different nodes so that if a machine fails the completion of the job does not collapse because blocks on the failed node have their copies on the other machines [8].

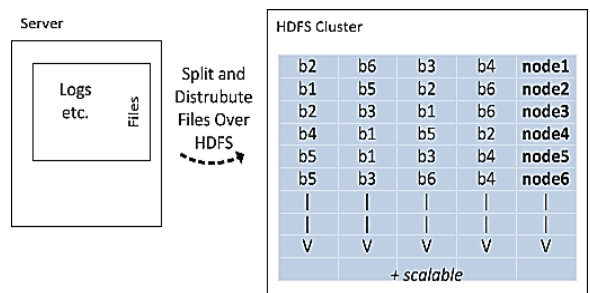

**Figure 2** Hadoop block structure

In a Hadoop cluster one machine works a master node and other nodes work as slaves. Master node has two roles which are Name Node and Job Tracker. Slave nodes have Data Node and Task Tracker roles. If a slave node fails and cannot finish executing its task, master node automatically schedules the same task to run on another slave machine [9].

The Hadoop platform created for this study can be shown as in Fig. 3.

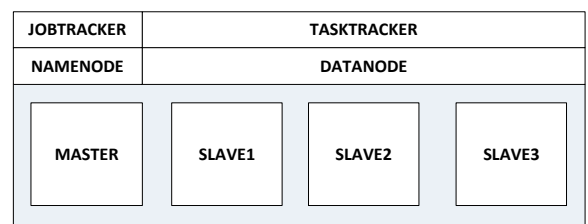

**Figure 3** Hadoop architecture

When setting up a Hadoop cluster Hadoop installations files are saved on a particular directory on the machines. The machines can be bare metal servers, virtual servers on a private cloud service from a paid provider. Cloud provider service can be infrastructure as a service or a service specific for Hadoop. Performance comparisons for these options have been done by academy and industry. There are some specific benchmarking works to show the advantages of different hardware sources. The platform used in this work is a server which has four cores of Intel 3.16GHz CPU and 16 GB of memory and uses Ubuntu 12.04 as the host operating system.

In this study we use OpenStack [10], an open-source cloud management platform which can be used as an Infrastructure as a Service software. One can easily set up and manage a cloud system with installing OpenStack on the host operating system.

Four instances of virtual machines were created as Hadoop cluster nodes. Fig. 4 shows the general architecture of the system.

Steps shown in Fig. 5 are common for all programs of this study. First input log files are split into blocks (64 MB each) to be distributed over the HDFS. Input path in the HDFS is given as a parameter to MapReduce job. MapReduce job starts with executing the map class over the distributed blocks. For text inputs map functions execute over each line of the input. So it is expected to have a list of map results as large as the number of lines of the input file.

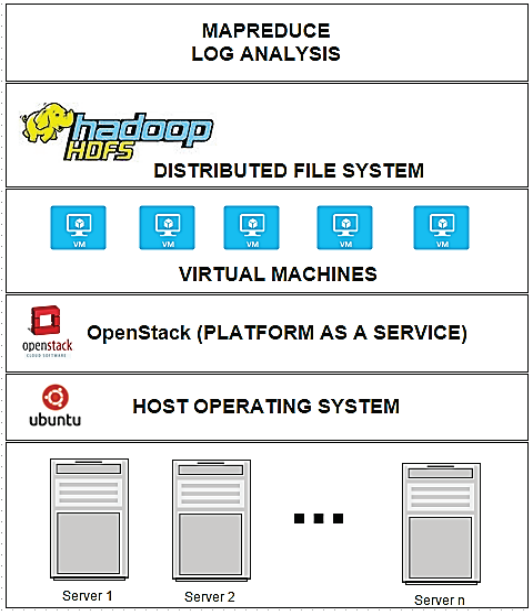

**Figure 4** OpenStack Cloud + Hadoop integration and architecture

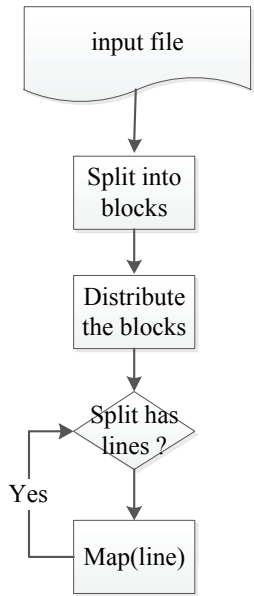

**Figure 5** Distributing the blocks and Map steps

The steps which are specific to problem start within the map function. In this study, extracting useful information from log files can be collected under two main titles.

1 - Single MapReduce job multiple outputs
2 - Chained MapReduce jobs for more complex scenarios

Format of the log files used in this study is as follows:
Date , Time , Client IP Address , User Name , Service Name and Instance Number , Server Name , Server IP, Address , Server Port , Method , URI Stem , URI Query , HTTP Status , Win32 Status , Bytes Sent , Bytes Received , Time Taken , Protocol Version , Host , User Agent , Cookie , Referrer , Protocol Substatus

Sample line:
2013-04-15    00:00:07    W3SVC1    10.1.1.5    GET /ilahiyat/Tr/BilimselFaaliyetler/FakulteDergisi.htm    -    80    -    66.249.78.66    Mozilla/5.0+(compatible;+Googlebot/2.1;++ http://www.google.com/bot.html) –www. firat.edu.tr

### 2.1 Single MapReduce job multiple outputs

Counting the occurrence of each word in each line gives us useful information like access count on a particular day or days or total number of requests made to a particular page etc. For this, each word in each line is mapped with value (1) and sum of the maps are grouped by their keys and they are printed out by the reducer.
Problem with this approach is that it is very hard to read from the output to find an answer for the questions such as given above. The solution we developed for this problem is writing multiple outputs such that each output gives different information. Single MapReduce job multiple output is a solution for this kind of problems.

In this approach map functions do mapping for each column in the log file by adding prefixes to the keys. Each key has a prefix to denote the file name for its results to be written. This operation is basically adding a prefix - which is a short string- to the beginning of the text. So that reducer can identify what that particular column key belongs to. For example value for day "2013-04-15" is emitted as (day_2013-04-15, 1), for page index.html it is (page_index.html, 1). The underscores here are used as delimiter for prefix names. So reducer iterates over the data and writes outputs conditionally to the key's prefixes. In this study one of the single MapReduce job multiple output tests resulted with the following files; browser_part-00000, ip_part-00000, day_part-00000, hour_part-00000, method_part-00000, page_part-00000. Separating results to files in this way makes it easy to view, analyze or visualize the results. Fig. 6 shows the structure of this method. Steps in Fig. 5 are expected to be performed before.

Coding behind these processes is done by using Hadoop API's MultipleOutputFormat abstract class which extends FileOutputFormat class. This class can be used by setting job configuration's set OutputFormat feature as MultipleOutputFormat class which in turn specifies the outputs.

By using this approach we obtain better run times for MapReduce jobs as opposed to running a different job for each column. Although there are same number of map and reduce operations in running a single job or in running multiple jobs (one job for each column) processes such as Job Initialization, Task Assignment, Progress and Status Updates are repeated redundantly for each job in the latter case.

### 2.2 Chained MapReduce jobs for more complex scenarios

MapReduce programming model provides a high performance solution for processing large volumes of data on modest computer clusters. Since MapReduce programs run in parallel, it should be considered that extra work is needed to be done when doing comprehensive analysis which requires running SQL like queries. Such a query usually requires running more than one MapReduce jobs. Few solutions have been proposed to deal with this problem. For example the HaLoop [11] framework is used to support iterative data processing workloads.

Rubao Lee and his friends developed a system [12] for optimizing sql like query jobs.

In this study we only use core Hadoop library for processing the queries. Tools like Pig, Hive, HBase were not used, instead we only focus on chaining jobs to solve complex scenarios.

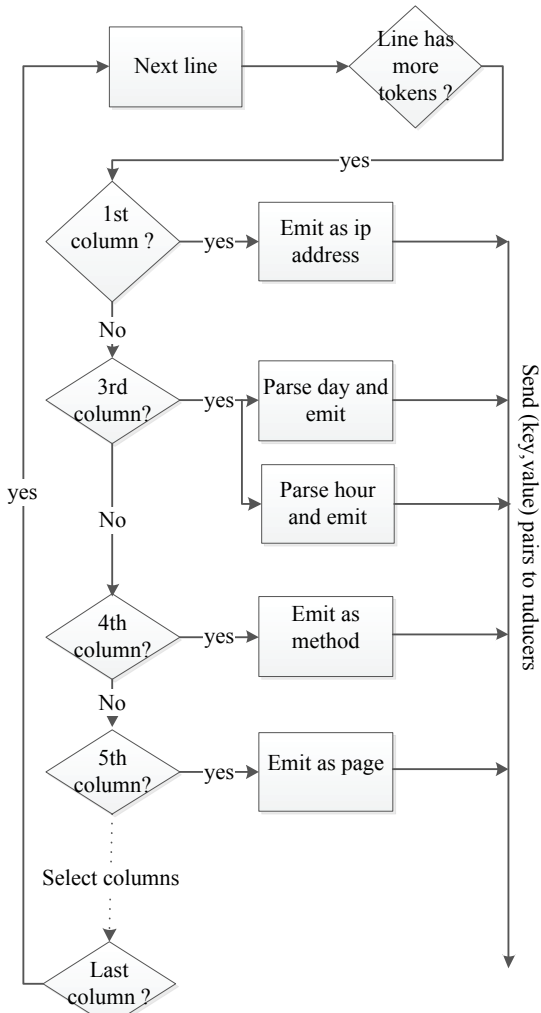

**Figure 6** MapReduce Job Structure

Here an example of job chaining is provided which deals with finding the total number of accesses to the most accessed pages in the busiest day of the web server. Assuming we know the busiest day, SQL equivalent of this query is:

```
SELECT pages FROM log
   WHERE day =
   SELECT MAX("day") FROM
   (
   SELECT COUNT(*) AS "day"
   FROM logs
   )
```

Three MapReduce jobs are needed to query this.
- Job1(Map1&Reduce1) for counting the visits on each day. Output1 resulted from this job which contains number of visits for each day.
- Job2(Map2&Reduce2) deals with sorting the result output1 to obtain the day which has the maximum value. Output2 resulted from this job gives the days in the output1 in ascending order. This job does not need a reduce operation but it should be used to invoke sorting operation between map and reduce phases. As a result we obtain the busiest day in terms of page accesses for the web server.
- Job3(Map3&Reduce3) deals with selecting the pages that are accessed on the busiest day which are obtained from Job2 so it is a conditional map operation and a reduce function. Output3 resulted from Job3 and it has the final output, hence the result of the SQL-like query.

Algorithm for this chained job is as follows: Firstly, output1 file should be obtained. This is done by a MapReduce job which consists of one map (Map1) and one reduce (Reduce1) processes. Output1 indicates days with their access frequencies, but for this problem only the day with highest frequency is needed. Hadoop does a sorting operation on the keys by default. As is explained in [2] it guarantees the intermediate key/value pairs to be in ascending order before they are sent to reducer. This ability can be used to sort days by their frequencies. Output1 has (day, total access) so its lines are split into token1=day and token2=access time and emitted as (token2, token1) and then these key/value pairs are sent to Reduce2. But before they are sent to reducer they are sorted by access time. Reduce2 is an identity reducer which only passes key/value pairs to output2. As a result output2 contains lines of total access/day pairs in increasing order. The last line of output2 indicates the busiest day which in turn parsed and assigned to a variable called max_day. This variable is passed to configuration object of the next job so that it can be accessed by mappers and reducers. Map3 lists accessed pages on max_day and Reduce3 sums the number of accesses of those particular pages.

## 3 Results

In this study we investigated the use and performance of popular distributed data processing approaches to web server log data. We show that Hadoop can help us process large amount of data where conventional data analysis approaches fall short. Web server logs have the tendency to grow exponentially to very big sizes and analyzing such big data requires high amount of resources in conventional approaches. In such cases Hadoop proves to be a feasible alternative with considerably low resource allocation requirements.

However, although Hadoop provides us with a high performance data processing alternative we still need several compute nodes for creating a Hadoop cluster since the real power of MapReduce can only be realized when the data are split across the distributed file system. Creating a cluster means running several instances of Hadoop on different machines. Virtualization helps us running several operating systems on a single physical machine which in turn can be used as Hadoop nodes. However since most virtualization software requires high license fees or extensive professional background we utilize an open source Cloud Computing software called OpenStack for creating the compute nodes of the Hadoop cluster.

**Table 1** Chained MapReduce jobs steps

|   | In | Emit | Out(file) |
|---|---|---|---|
| 1. | **MAP1**(line number, line) | **EMIT**(day, one) | |
| 2. | **REDUCE1**(day, [1,1,1…n]) | **EMIT**(day, $\sum_{i=0}^{i \leq n} value(i)$) | = output1 (day, total access) |
| 3. | **MAP2**(line number, line) | **EMIT**(token2, token1) | |
| 4. | **REDUCE2**(total access, day) | **EMIT**(total access, day) | = output2 (total access, day) |
| 5 | Set max_day variable as configuration parameter with the maximum accessed day | | |
| 6. | **MAP3**(line number, line) | if(day="max_day") Emit(page, one) | |
| 7. | **REDUCE3**(page,[1,1,1…n]) | **EMIT**(page, $\sum_{i=0}^{i \leq n} value(i)$) | = output3 (page, total access) |

Setting up Hadoop clusters manually can be very complex and daunting. Enterprise software such as Cloudera, HortonWorks and MapR can automate setting up and managing such clusters.

The study shows that if the analysis is being performed on a particular field on the file which is also independent from other fields then the analysis can be carried out relatively straight forward.

If the analysis requires results from the relationships of different fields which cannot be computed in parallel it can still be done easily. However in such cases several (sometimes tens even hundreds of) analyses (jobs) are carried out and the results from some are used as input to others. Which in turn requires running a job after some particular job is completed. In such cases job chaining can be used.

In this study we use job chaining for complex scenarios such as SQL-like queries. Output files in HDFS are used as input files for the next job, or the results from the output files are used as input parameters for the next job.

Fig. 7 shows analysis time over the data size.

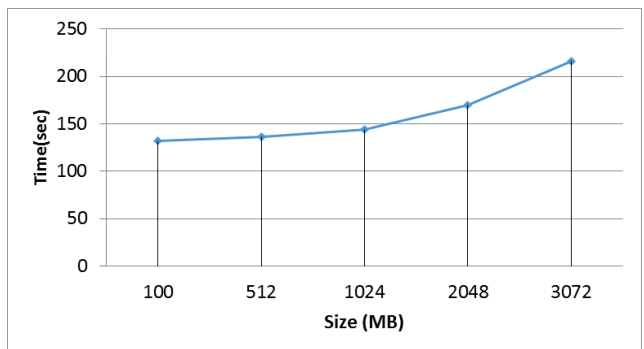

**Figure 7** Hadoop analysis time

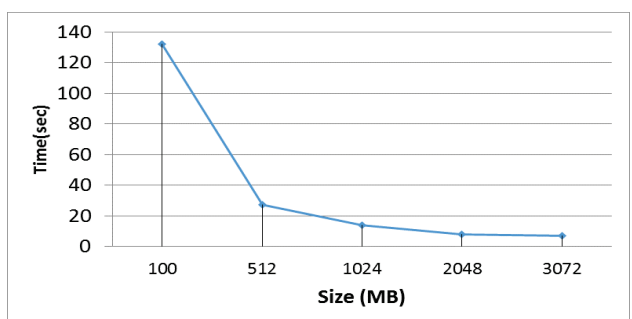

**Figure 8** Hadoop performance

In Fig. 8, time indicates analysis time per 100 MB of data for the purpose of comparing the running times for the same amount of data. Mathematically time values are calculated by (running_time)*(100/size) equation. As it can be seen from the figure performance of data processing grows exponentially as the data increases. We expect that the increase of the performance would be much higher if the size of the data was bigger.

## 4 Performance comparison with Spark

Analyzing large data sets has become a very important subject for the industry and research world. Parallel data processing frameworks are becoming faster by using new methods. However this fashion has started with Google's open source implementation Hadoop and its ecosystem members such as Hive and Pig SQL processing engines, the game is still on. Hadoop is mainly optimized for using disk data and jobs which complete in one iteration. But analysis of big data usually needs complex, multi-iteration processes [13].

Spark is one of the latest framework in Apache Big Data Stack. Spark suggests that it is 10× faster than Hadoop [14]. Unlike Hadoop, Spark supports in-memory processing and this becomes very useful in iterative computing of big data. Spark comes with different cluster manager options such as Amazon EC2, Standalone, Mesos and Yarn [15].

In this study an Apache Spark cluster was set up in Standalone Deploy Mode. Spark version 0.9 was installed on virtual machines. Spark uses the RDDs (Resilient Distributed Datasets) data type. RDDs can be constructed from HDFS [15]. This means a log directory on a distributed file system can be processed by Spark as well as Hadoop without any additional work done.

Spark's performance over Hadoop is investigated with word count scenarios. According to results obtained in this study Spark runs faster than Hadoop even in the first iteration.

A program, searching a specific word in Apache Log files is implemented in MapReduce Java and Spark Java. Hadoop and Spark were installed on the instances of the cluster. Running times are shown in Fig. 9 (Hadoop vs Spark 10 instances) and Fig. 10 (Hadoop vs Spark 4 instances). Running times in the graphics are obtained with 1 iteration.

It has been observed that Spark performs closer to Hadoop as the data size grows. This is because Spark which works in-memory, starts to work in disk I/O like Hadoop when the size of input data exceeds the memory capacity. However this is the case for the processes which consist of one iteration; when the number of iterations increases Spark can perform 10 times better because the data would already be in the memory.

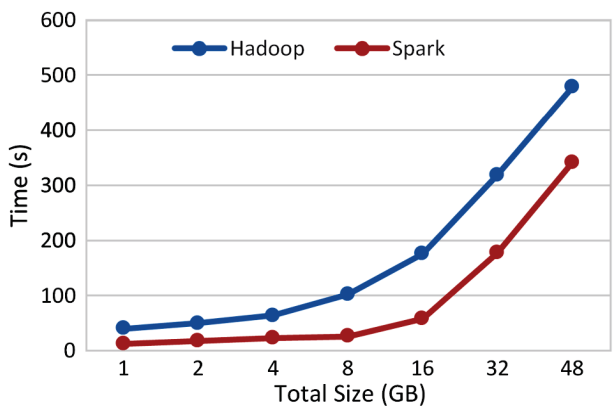

**Figure 9** Hadoop vs Spark (10 instances)

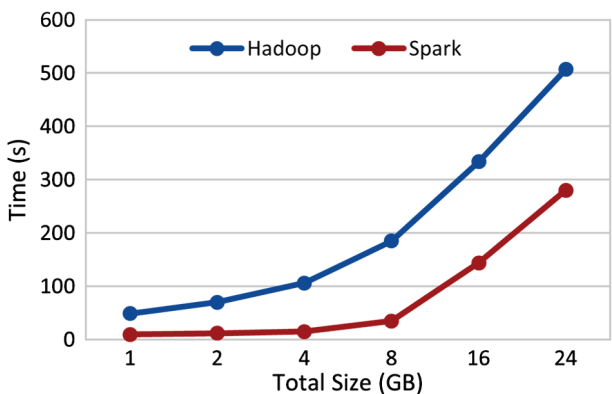

**Figure 10** Hadoop vs Spark (4 instances)

Ref. [15] proves this with Logistic Regression benchmark. In [15] there is performance comparison of Hadoop vs Spark; they run logistic regression algorithm for 29 GB of data. Running time of each iteration on Hadoop takes 127 s and first iteration on Spark takes 174 s. But after the first iteration, subsequent iterations in Spark take 6 s.

Additionally we observed that Spark's interactive shell feature is a very useful tool for log analysis, operations like searching, counting etc. The functional programming interface provided by Spark's shell makes it possible to use a distributed file called RDDs as an input and do parallel operations on the data without writing a full program code. Simple commands in Scala like logFile.filter(line => line.contains("error")).count() //how many lines contain error make Spark a practical tool for log analysis.

## 5 Conclusion

In this paper we proposed a cloud-based Big Data Analysis system for analyzing large amounts of log data, specifically web server logs. The system can automatically detect and analyze several types of web server logs. We used open source Cloud Computing and Big Data technologies for setting up the system. The performance evaluations show that the system can be used to analyze very large log files in a reasonable amount of times. We also present a comparison of the Hadoop based system with more recent Spark framework.

Authors' addresses

*Galip Aydin, PhD.*
Firat University, Faculty of Engineering,
Computer Engineering Department,
Elazig, Turkey
Contact Tel.: +90 4242370000-3150
E-mail: gaydin@firat.edu.tr

*Ibrahim R. Hallac, MSc.*
Firat University, Faculty of Engineering,
Computer Engineering Department,
Elazig, Turkey
Contact Tel.: +90 4242370000-6303
E-mail: irhallac@firat.edu.tr